\documentclass[12pt]{article}
\input xy
\xyoption{all}
\input epsf.tex
\usepackage{pstricks}
\usepackage{psfrag}

\usepackage{graphicx}
\usepackage[small,loose]{subfigure}
\usepackage{mathrsfs}
\usepackage{bbm} 
\usepackage{cancel}
\usepackage{amssymb,amsmath}

\def\harr#1#2{\smash{\mathop{\hbox to .3in{\rightarrowfill}}
 \limits^{\scriptstyle#1}_{\scriptstyle#2}}}

\def\s2{\frac{1}{\sqrt2}}

\def\be{\begin{equation}}
\def\ee{\end{equation}}
\def\beqa{\begin{eqnarray}}
\def\eeqa{\end{eqnarray}}

\def\Dsl{\,\raise.15ex\hbox{/}\mkern-13.5mu D} 
\def\d3{d^3}

\newcommand{\dif}{\mbox{d}}
\newcommand{\im}{\mbox{i}}

\begin{document}

\vspace{.5cm}
\begin{center}
\Large{\bf On the Moyal deformation of Nahm Equations in seven dimensions}\footnote{This paper is devoted
to the memory of Prof. Guillermo Moreno Rodr\'{\i}guez.}\\
\vspace{1cm}

\large Hugo Garc\'{\i}a-Compe\'an\footnote{e-mail address: {\tt
compean@fis.cinvestav.mx}}, Aldo A.
Mart\'{\i}nez-Merino\footnote{e-mail address: {\tt
amerino@fis.cinvestav.mx}}
\\
[2mm] {\small \em Departamento de F\'{\i}sica, Centro de
Investigaci\'on y de
Estudios Avanzados del IPN}\\
{\small\em P.O. Box 14-740, 07000 M\'exico D.F., M\'exico}
\\[4mm]

\vspace*{2cm}
\small{\bf Abstract} \\
\end{center}

\begin{center}
\begin{minipage}[h]{14.0cm}
{ We show how the reduced (anti-)self-dual Yang-Mills equations in seven dimensions described by the Nahm equations can be carried over to the Weyl-Wigner-Moyal formalism. In the process some new solutions for the cases of gauge groups $\mbox{SU}(2)$ and $\mbox{SL}(2, \mathbb{R})$ are explicitly obtained. }
\end{minipage}
\end{center}

\bigskip
\bigskip

\date{\today}

\vspace{3cm}

\leftline{August, 2009}

\newpage

\section{Introduction}

Self-dual Yang-Mills equations in higher dimensions have been studied since many years ago
\cite{CDFN,Ward:1983zm}. Moreover these equations have been used to study Yang-Mills instantons, particularly in eight-dimensions (see for instance, \cite{figueroa}). It has been interesting also from the pure mathematical point of view \cite{Donaldson:1996kp}. There is also ADHM version for the construction of Yang-Mills instantons in eigh-dimansions \cite{Hiraoka:2002wm,Buniy:2002jw}.
Since the early steps have been clear that the structure involves the idea of octonion algebra associated to the eight dimensional manifolds of Spin(7)-holonomy and $G_2$ for the seven dimensional case (for some reviews on special holonomy manifolds see \cite{H&L,Joyce,Salamon}). In fact they have been used in perturbative and non-perturbative compactifications of string theory \cite{HS,Shatashvili:1994zw,Acharya:2004qe}. More recently it was studied how to construct a cohomological field theory on higher dimensions \cite{Baulieu:1997jx,Acharya:1997jn}. The study of self-dual metrics and gravitational instantons in eight-dimensional spaces have been considered in Refs. \cite{Acharya:1996tw,Floratos:1998ba,Bakas:1998rt}.

On the other hand, dimensional reductions of the self-dual Yang-Mills equations in eight dimensions to
lower dimensions has been explored. In particular, to seven dimensions on manifolds of $G_2$ holonomy gives rise to the seven-dimensional version of the Nahm equations, a bit motivated by the M(atrix) Model proposal \cite{Banks:1996vh}.
Nahm equations in seven dimensions have been studied in \cite{Curtright:1997st,Fairlie:1997vj}. The solutions to the top equations were examined in Refs. \cite{FU,Ueno}. Moreover the search of an abelian structure mimicking Seiberg-Witten equations and the effect of S-duality was discussed in \cite{GT}. Later Baker and Fairlie in \cite{Baker:1999ru} constructed a large class of matrix representations of the algebra of residues of the variables at simple poles. In the present paper we will use these solutions to the find new solutions to a Moyal deformation of Nahm's equations in seven dimensions. Moreover in \cite{Baker:1999ru} was taken the large-$N$ limit by considering the Moyal-Nahm equations in the seven-dimensional case.
After that they obtain a set of non-trivial solutions to these equations. In this paper we will derive the Moyal-Nahm equations from the Weyl-Wigner-Moyal (WWM) formalism that we will explain later.

On the other hand it is well known that in four dimensions self-dual metrics can be described in terms of
a self-dual Yang-Mills theory (and also full Yang-Mills theory) whose connections are volume preserving-valued
vector fields. These vector fields satisfy the Ashtekar-Jacobson-Smolin (AJS) equations \cite{ajs}. Thus,
through an appropriate large-$N$ limit of self-dual Yang-Mills theories and their dimensional reductions is possible to find self-dual metric by solving the resulting equations \cite{Castro1,Husain,Castro2,Compeans}.

One would ask if there is some suitable large $N$-limit for the Yang-Mills equations in eight dimensions such that one can extract a gravitational instanton solution. A recent proposal has been
that of Ref. \cite{yo}, where a set of equations has been proposed that generalizes AJS equations to eight-dimensions. In this same context there were constructed some solutions of metrics with Spin(7) and $G_2$ holonomy \cite{KN}. A proposal in the direction of generalizing  the Ashtekar's formalism to eight dimensions is discussed in Ref. \cite{nieto}. In spite of the formulation of these proposals the subject is still unconcluded. In the present paper we studied a system that would be the first step to address this problem. Thus we studied the large-$N$ limit of the Nahm equations in seven dimensions by taking first its Moyal deformation. To do that we take the WWM formalism in a similar spirit to \cite{Compean1}, we make a WWM-description of the Nahm equations in seven dimensions. In the
$\hbar \to 0$ limit we shall get their large-$N$ limit. In self-dual Yamg-Mills in four dimensions it is known that large $N$-limit of Nahm equations gives an hyper-Kahler (self-dual) metric of the underlying four dimensional spacetime \cite{Compean1}.

A natural question is whether these equations are integrable trying to imitate \cite{wardlargeN}. Another issue is the possibility to address the problem of the integrability of the large-$N$ limit in the full Yang-Mills and the full Einstein equations in eight dimensions trying to imitating the twistor construction as suggested in \cite{wittenlargeN}. In this paper we will see the effect of the large-N limit in systems in eight-dimensional manifolds of Spin(7) holonomy.

The paper is organized as follows. In Section 2 we give an overview on the
octonions in order to fix the notation, conventions and the properties that will be
useful later.
Section 3 is devoted to review the self-dual Yang-Mills equations in eight dimensions.
In this section we also recall the reduction of the self-dual equations to seven dimensions.
Particularly we recall the structure of the Nahm equations. In Section 4 we proceed to study the
WWM deformation of the Nahm equations in seven dimensions. Then in Sec. 5 we found some new solutions
of the Moyal deformation of the Nahm equations by a systematic method of the WWM formalism used already in another systems. Finally, in Sec. 6 we give our final remarks.

\section{Octonionic Preliminaries}

 In the present section we overview some basic material about octonion algebra, in order to fix notation, conventions  and results for future reference (for more details see, for instance, \cite{H&L,PMS,JBaez,DGTze}). The octonions $\mathbb{O}$ are one of the four real division algebras, among  the real numbers, $\mathbb{R}$, the complex numbers, $\mathbb{C}$, and quaternions, $\mathbb{H}$. They consist of seven imaginary units $e_{1}, \dots, e_{7}$, and one real, $e_{0} = 1$, that satisfy
\begin{equation}
e_{i} e_{j} = C_{ijk} e_{k} - \delta_{ij}, \quad \mbox{for} \quad i, j, k = 1, \dots, 7,
\end{equation}

\noindent where the quantity $C_{ijk}$ is a totally antisymmetric tensor defined by
\begin{equation}
C_{ijk} = +1 \quad \mbox{for} \quad ijk = 123, 617, 257, 536, 145, 246, 347.
\end{equation}

\noindent Such a tensor have its dual, being the four index quantity which is defined by
\begin{equation}
C_{ijkl} = \frac{1}{3!} \varepsilon_{ijklmno} C_{mno},
\end{equation}
where, as in the case of $C_{ijk}$, its values are given by the combinations
\begin{equation}
C_{ijkl} = +1, \quad \mbox{for} \quad ijkl = 4567, 4532, 1463, 1427, 2367, 1357, 1256.
\end{equation}

Because of the non-associativity of $\mathbb{O}$, the $C_{ijk}$ and its dual, do not satisfy the usual relation involving the Levi-Civita antisymmetric symbol. That is, they satisfy
\begin{eqnarray}
C_{ijk}C_{mnk} & = & \delta_{im} \delta_{jn} - \delta_{in} \delta_{jm} - C_{ijmn}.
\label{EQ1}
\end{eqnarray}

Also, again by the non-associativity, they do not satisfy the Jacobi identity, they instead satisfy a relation known as the {\it Moufang identity} \cite{PMS}, which is
\begin{equation}
[x, [y, z]] + [y, [z, x]] + [z, [x, y]] = 2 [x, y, z],
\end{equation}
where the quantity $[x, y, z]$ for all $x, y, z \in \mathbb{O}$, is the so called {\it associator}, defined by
\begin{equation}
[x, y, z] = (x y) z - x (y z),
\end{equation}
which, like the commutator, it measures at what extent some octonion triple do not associate.

The elements $C_{ijk}$ defines a 3-form $\omega$ which is invariant under the action of the exceptional Lie group $G_{2}$. The 3-form $\omega$ together with its dual defines a Riemannian manifold which has its holonomy group contained in the mentioned Lie group \cite{Salamon, Joyce}. Explicitly, the 3-form has the form
\begin{equation}
\omega = \omega_{123} - \omega_{167} - \omega_{527} - \omega_{563} - \omega_{154} - \omega_{264} - \omega_{374},
\end{equation}
where we are using the notation $\omega^{ijk} = f^{i} \wedge f^{j} \wedge f^{k}$, being $f^{1}, \dots, f^{7}$ the dual basis to the $e$'s.

With the 3-form $\omega$ it is possible to define a four form $\Omega$ which is invariant under $\mbox{Spin}(7)$, which is termed the {\it Cayley form}. If we consider $e_{0}^{*}$ as the dual to $e_{0}$, $\Omega$ is defined by
\begin{equation}
\Omega = e_{0}^{*} \wedge \omega + *_{7} \omega,
\label{Cayleyform}
\end{equation}

\noindent where the second term, $*_{7} \omega$, is the Hodge dual form in seven dimensions of $\omega$. This form defines a Riemannian manifold with its holonomy group being contained in $\mbox{Spin}(7)$.\\

\section{The Self-Dual Equations}

One of the main problems in the theory of gauge fields, is to address the existence of solutions of Yang-Mills equations in higher dimensions. Another important question concerns the integrability of these equations. Remember that the integrability of full Yang-Mills in four dimensions remains elusive until now. However we know the complete answer when the (anti-)self-dual conditions are satisfied. The solutions to these equations are the so called {\it instantons}, which are solutions to the self-dual Yang-Mills equations of finite action. In \cite{CDFN}, the authors studied the possible groups under which a four antisymmetric tensor would be invariant in dimensions greater than four. That is, starting from the ``self-dual equation''
\begin{equation}
F_{\alpha \beta} = \lambda T_{\alpha \beta \gamma \delta} F^{\gamma \delta},
\end{equation}
where the quantity $T_{\alpha \beta \gamma \delta}$ is totally anti-symmetric, one would wonder what groups $G$  leave $T_{\alpha \beta \gamma \delta}$ (transforming irreducibly under $G$) invariant. Here $F_{\alpha \beta} = \partial_{\alpha} A_{\beta} - \partial_{\beta} A_{\alpha} + [A_{\alpha}, A_{\beta}]$ represents the curvature two-form of the gauge connection $A$ in the given manifold. The cases discussed in that paper \cite{CDFN}, ranges from $n={\rm dim}(M)=4,5,6,7,8,9$. Here we will be interested in the case $n=8$.

Consider a Riemannian manifold, $M$, in eight dimensions whose holonomy group lies in $\mbox{Spin}(7)$. This manifold is defined by the existence of a four-form which indeed is invariant under $\mbox{Spin}(7)$. Such a four form is precisely the Cayley form (\ref{Cayleyform}). Thus, we have that the self-duality equation in eight dimensions is
\begin{equation}
F_{\alpha \beta} = \lambda \Omega_{\alpha \beta \gamma \delta} F^{\gamma \delta},
\end{equation}
for which we have that the eigenvalue $\lambda$ take the values of $1$ and $-3$. From now on, the Greek indices runs from 0 to 7. Then, we can write down  this last equation for the self-dual case ($\lambda =1$) as
\begin{equation}
F = *_{8} (\Omega \wedge F).
\label{selfduality}
\end{equation}
These are seven equations, explicitly written in the form
\begin{eqnarray}
F_{01} &=& F_{23} + F_{76} + F_{45}, \nonumber \\
F_{02} &=& F_{31} + F_{57} + F_{46}, \nonumber \\
F_{03} &=& F_{12} + F_{65} + F_{47}, \nonumber \\
F_{04} &=& F_{51} + F_{62} + F_{73}, \label{SDEqs} \\
F_{05} &=& F_{72} + F_{36} + F_{14}, \nonumber \\
F_{06} &=& F_{17} + F_{53} + F_{24}, \nonumber \\
F_{07} &=& F_{61} + F_{25} + F_{34}. \nonumber
\end{eqnarray}
These equations take a succinct form remembering the fact that $\Omega_{0ijk} = C_{ijk}$, and consequently
\begin{equation}
F_{0i} = \frac{1}{2} C_{ijk} F_{jk}.
\end{equation}
The other equations with the eigenvalue of $-3$ are 21 of them, and they will not be considered in the present paper.

In order to reduce the equations to lower dimensions, we choose a gauge in which the component $A_{0}$ is zero by considering the static case, where all the other components of the connection depends only on one variable $t$. Thus equations (\ref{SDEqs}) take the form (taking into account that with such a selection of coordinates, we could consider $e_{0}^{*} = \dif t$)
\begin{eqnarray}
\frac{\dif A_{1}}{\dif t} &=& [A_{2}, A_{3}] + [A_{7}, A_{6}] + [A_{4}, A_{5}], \nonumber \\
\frac{\dif A_{2}}{\dif t} &=& [A_{3}, A_{1}] + [A_{5}, A_{7}] + [A_{4}, A_{6}], \nonumber \\
\frac{\dif A_{3}}{\dif t} &=& [A_{1}, A_{2}] + [A_{6}, A_{5}] + [A_{4}, A_{7}], \nonumber \\
\frac{\dif A_{4}}{\dif t} &=& [A_{5}, A_{1}] + [A_{6}, A_{2}] + [A_{7}, A_{3}], \\
\frac{\dif A_{5}}{\dif t} &=& [A_{7}, A_{2}] + [A_{3}, A_{6}] + [A_{1}, A_{4}], \nonumber \\
\frac{\dif A_{6}}{\dif t} &=& [A_{1}, A_{7}] + [A_{5}, A_{3}] + [A_{2}, A_{4}], \nonumber \\
\frac{\dif A_{7}}{\dif t} &=& [A_{6}, A_{1}] + [A_{2}, A_{5}] + [A_{3}, A_{4}]. \nonumber
\end{eqnarray}

Again, the equations could be rewritten in the form
\begin{equation}
\frac{\dif A_{i}}{\dif t} = \frac{1}{2} C_{ijk} [A_{j}, A_{k}],
\label{Nahm7}
\end{equation}
and these are the {\it Nahm Equations} in seven dimensions.\\

\noindent {\it Remark}. We note that the same equations were considered in Ref. \cite{Baker:1999ru}, but different conventions were taken, i.e., a different octonion multiplication table. But considering that the automorphisms group of the imaginary part of $\mathbb{O}$ is $G_{2}$, a symmetry of the equations under this group is supposed to be fulfilled.

At this stage, we have not considered yet the gauge group where the connection take its values. Let ${\bf G}$ be the gauge group with Lie algebra $\mathcal{G}$ which we assume to be semi-simple, and therefore $A \in \mathfrak{M} \otimes \mathcal{G}$. Here $\mathfrak{M}$ is the set of seven $(6 \times 6)$-matrices $B_i=C_{ijk}$ constructed with $C_{ijk}$ \cite{Baker:1999ru}. Moreover we consider $\mbox{Tr}(\cdot)$ to be the invariant bilinear form defined in $\mathfrak{M}\otimes \mathcal{G}$. Using the equation (\ref{Nahm7}), we see that the Gauss law is satisfied, taking it as a constraint of the theory, which is
\begin{equation}
\Big[ \frac{\dif A_{i}}{\dif t}, A_{i} \Big] = 0.
\label{GaussLaw}
\end{equation}

Taking a second derivative on $A_{i}$ by using (\ref{Nahm7}), we obtain the equation of motion of the gauge field
\begin{equation}
\frac{\dif^2 A_{i}}{\dif t^2} + [A_{j}, [A_{j}, A_{i}]] + \frac{1}{2}C_{ijkl}[A_{j}, [A_{k}, A_{l}]] = 0. 
\label{EqMotion}
\end{equation}
We consider these equations as an extension of the Nahm's equations of motion in three dimensions. The variational principle from which we get this equation of motion comes from the Lagrangian given by the next expression
\begin{equation}
\mathcal{L} = \alpha \mbox{Tr} \Big\lbrace \frac{\dif A_{i}}{\dif t} \frac{\dif A_{i}}{\dif t} + \frac{1}{2} [A_{i}, A_{j}][A_{i}, A_{j}] - \frac{1}{4}C_{ijkl}[A_{i}, A_{j}][A_{k}, A_{l}] \Big\rbrace,
\end{equation}
where $\alpha$ is a constant. From this Lagrangian, immediately we write down the energy function associated with the theory
\begin{equation}
\mathcal{E} = \mbox{Tr} \Big\lbrace \frac{\dif A_{i}}{\dif t} \frac{\dif A_{i}}{\dif t} - \frac{1}{2} [A_{i}, A_{j}][A_{i}, A_{j}] + \frac{1}{4}C_{ijkl}[A_{i}, A_{j}][A_{k}, A_{l}] \Big\rbrace,
\end{equation}
which is zero when we assume that $A_i$ is a solution to Eqs. (\ref{Nahm7}). One can check easily that ${\cal E}$ is not positive definite.

\section{The Moyal Deformation of the Nahm Equations in Seven Dimensions}

In this section we formulate the Moyal deformation of the theory discussed in the previous section. In first place, we regard the components of the gauge field as operator-valued quantities, acting on a Hilbert space $\mathcal{H} = L^{2}(\mathbb{R})$. Here we choose $| \phi_{n} \rangle$ with $n = 0, 1, \dots$ as an orthonormal basis of $\mathcal{H}$. Such a basis is complete and satisfy a closure relation
\begin{equation}
\langle \phi_{n} | \phi_{m} \rangle = \delta_{nm}, \quad \sum | \phi_{n} \rangle \langle \phi_{n} | = \widehat{I},
\end{equation}
where $\widehat{I}$ is the identity operator. Then we will get the correspondence: $A_{i} \rightarrow \widehat{A}_{i} \in \mathfrak{M} \otimes \widehat{\mathcal{U}}$, with $\widehat{\mathcal{U}}$ being the Lie Algebra of anti-self-dual operators acting on $L^{2}(\mathbb{R})$. Now we promote the equations (\ref{EqMotion}) and (\ref{GaussLaw}) to its operator version by making this change in the $A$'s and taking the commutator instead of the Lie brackets, $[\cdot , \cdot ]$. Thus
\begin{equation}
\frac{\dif^2 \widehat{A}_{i}}{\dif t^2} + [\widehat{A}_{j}, [\widehat{A}_{j}, \widehat{A}_{i}]] + \frac{1}{2}C_{ijkl}[\widehat{A}_{j}, [\widehat{A}_{k}, \widehat{A}_{l}]] = 0, \label{qEqMotion}
\end{equation}
and
\begin{equation}
\Big[ \frac{\dif \widehat{A}_{i}}{\dif t}, \widehat{A}_{i} \Big] = 0. \label{qGaussLaw}
\end{equation}
In order to apply the WWM formalism to these equations, we make a redefinition of the connection $\widehat{A}$ to quantities where the deformation parameter $\hbar$ is explicitly given. Such definition is
\begin{equation}
\widehat{\mathcal{A}}_{i} := \im \hbar \widehat{A}_{i},
\end{equation}
and in consequence, the equations (\ref{qEqMotion}) and (\ref{qGaussLaw}) are rewritten as
\begin{equation}
\frac{\dif^2 \widehat{\mathcal{A}}_{i}}{\dif t^2} + \frac{1}{\im \hbar}\Big[ \widehat{\mathcal{A}}_{j}, \frac{1}{\im \hbar}\Big[ \widehat{\mathcal{A}}_{j}, \widehat{\mathcal{A}}_{i} \Big] \Big] + \frac{1}{2\im \hbar}C_{ijkl} \Big[ \widehat{\mathcal{A}}_{j}, \frac{1}{\im \hbar} \Big[ \widehat{\mathcal{A}}_{k}, \widehat{\mathcal{A}}_{l} \Big] \Big] = 0, \label{QEqMotion}
\end{equation}
and the Gauss Law is
\begin{equation}
\frac{1}{(\im \hbar)^2} \Big[ \frac{\dif \widehat{\mathcal{A}}_{i}}{\dif t}, \widehat{\mathcal{A}}_{i} \Big] = 0. \label{QGaussLaw}
\end{equation}

For this operator version, the Lagrangian and the Hamiltonian giving rise to equations (\ref{QEqMotion}) is
\begin{eqnarray}
\mathcal{L}^{(q)} &=& \mbox{Tr} \Big[ 2\pi \hbar \Big\lbrace \frac{\dif \widehat{\mathcal{A}}_{i}}{\dif t} \frac{\dif \widehat{\mathcal{A}}_{i}}{\dif t} - \frac{1}{2\hbar^{2}} [\widehat{\mathcal{A}}_{i}, \widehat{\mathcal{A}}_{j}][\widehat{\mathcal{A}}_{i}, \widehat{\mathcal{A}}_{j}] + \frac{1}{4\hbar^{2}}C_{ijkl}[\widehat{\mathcal{A}}_{i}, \widehat{\mathcal{A}}_{j}][\widehat{\mathcal{A}}_{k}, \widehat{\mathcal{A}}_{l}] \Big\rbrace \Big] \nonumber \\
&=& 2\pi \hbar \sum_{n} \Big\langle \phi_{n} \Big| \frac{\dif \widehat{\mathcal{A}}_{i}}{\dif t} \frac{\dif \widehat{\mathcal{A}}_{i}}{\dif t} - \frac{1}{2\hbar^{2}} [\widehat{\mathcal{A}}_{i}, \widehat{\mathcal{A}}_{j}][\widehat{\mathcal{A}}_{i}, \widehat{\mathcal{A}}_{j}] + \frac{1}{4\hbar^{2}}C_{ijkl}[\widehat{\mathcal{A}}_{i}, \widehat{\mathcal{A}}_{j}][\widehat{\mathcal{A}}_{k}, \widehat{\mathcal{A}}_{l}] \Big| \phi_{n} \Big\rangle \nonumber \\
& &
\end{eqnarray}
and
\begin{eqnarray}
\mathcal{E}^{(q)} &=& \mbox{Tr} \Big[ 2\pi \hbar \Big\lbrace \frac{\dif \widehat{\mathcal{A}}_{i}}{\dif t} \frac{\dif \widehat{\mathcal{A}}_{i}}{\dif t} + \frac{1}{2\hbar^{2}} [\widehat{\mathcal{A}}_{i}, \widehat{\mathcal{A}}_{j}][\widehat{\mathcal{A}}_{i}, \widehat{\mathcal{A}}_{j}] - \frac{1}{4\hbar^{2}}C_{ijkl}[\widehat{\mathcal{A}}_{i}, \widehat{\mathcal{A}}_{j}][\widehat{\mathcal{A}}_{k}, \widehat{\mathcal{A}}_{l}] \Big\rbrace \Big] \nonumber \\
&=& 2\pi \hbar \sum_{n} \Big\langle \phi_{n} \Big| \frac{\dif \widehat{\mathcal{A}}_{i}}{\dif t} \frac{\dif \widehat{\mathcal{A}}_{i}}{\dif t} + \frac{1}{2\hbar^{2}} [\widehat{\mathcal{A}}_{i}, \widehat{\mathcal{A}}_{j}][\widehat{\mathcal{A}}_{i}, \widehat{\mathcal{A}}_{j}] - \frac{1}{4\hbar^{2}}C_{ijkl}[\widehat{\mathcal{A}}_{i}, \widehat{\mathcal{A}}_{j}][\widehat{\mathcal{A}}_{k}, \widehat{\mathcal{A}}_{l}] \Big| \phi_{n} \Big\rangle. \nonumber \\
& &
\end{eqnarray}

Let $\mathcal{B}$ and $C^{\infty}(\Sigma, \mathbb{R})$ be the sets of self-adjoint linear operators acting on the Hilbert space $\mathcal{H} = L^{2}(\mathbb{R})$ and the space of infinite defferentiable functions defined on the phase space $\Sigma$, respectively. The {\it Weyl correspondence}, $\mathcal{W}^{-1}$, is a one-one correspondence between these two sets, $\mathcal{W}^{-1}: \mathcal{B} \rightarrow C^{\infty}(\Sigma, \mathbb{R})$, given by
\begin{eqnarray}
\mathcal{A}^{j}(t, p, q; \hbar) &\equiv& \mathcal{W}^{-1}(\widehat{\mathcal{A}}^{j}) \nonumber \\
&:=& \int_{-\infty}^{\infty} \Big\langle q - \frac{1}{2}\xi \Big| \widehat{\mathcal{A}}^{j}(t) \Big| q + \frac{1}{2}\xi \Big\rangle \exp\left(\frac{\im}{\hbar} \xi \cdot p\right) \dif \xi,
\end{eqnarray}
for all $\widehat{\mathcal{A}} \in \mathcal{B}$ and $\mathcal{A} \in C^{\infty}(\Sigma, \mathbb{R})$. Thus, the operator version of Nahm equations in this situation is
\begin{equation}
\frac{\dif \widehat{\mathcal{A}}_{i}}{\dif t} = \frac{1}{2i\hbar} C_{ijk} [\widehat{\mathcal{A}}_{j}, \widehat{\mathcal{A}}_{k}].
\end{equation}

As is standard, we define the Moyal product $\star$ over the set of functions $C^{\infty}(\Sigma, \mathbb{R})$, by the relation
\begin{equation}
\mathcal{F}_{i} \star \mathcal{F}_{j} := \mathcal{F}_{i} \exp \left( \frac{\im \hbar}{2} \stackrel{\leftrightarrow}{\mathcal{P}} \right) \mathcal{F}_{j},
\end{equation}
where $\stackrel{\leftrightarrow}{\mathcal{P}}$ is the operator
\begin{equation}
\stackrel{\leftrightarrow}{\mathcal{P}} := \frac{\stackrel{\leftarrow}{\partial}}{\partial q} \frac{\stackrel{\rightarrow}{\partial}}{\partial p} - \frac{\stackrel{\leftarrow}{\partial}}{\partial p} \frac{\stackrel{\rightarrow}{\partial}}{\partial q},
\end{equation}
and $\mathcal{F}_{j} = \mathcal{F}_{j} (t, p, q; \hbar)$. The arrows stand for over what functions the partial derivatives will act. As usual we note that for $\hbar \rightarrow 0$ we recover the usual product between functions and we get the large $N$-limit su$(\infty)$ gauge theory.

Using the Weyl correspondence, the bilinear operator $[\cdot, \cdot]$ has its deformed version, which is
\begin{equation}
\mathcal{W}^{-1} \left(\frac{1}{\im \hbar} [\widehat{\cal F}_{i}, \widehat{\cal F}_{j} ]\right) = \frac{1}{\im \hbar} (\mathcal{F}_{i} \star \mathcal{F}_{j} - \mathcal{F}_{j} \star \mathcal{F}_{i}) := \{\mathcal{F}_{i}, \mathcal{F}_{j} \}_{M},
\end{equation}
where $\{\cdot ,\cdot \}_{M}$ stands for the {\it Moyal bracket}. In the limit $\hbar \to 0$ Moyal bracket reduces to that of Poisson. The Weyl correspondence establishes an isomorphism between the algebra $(\mathcal{B}, [\cdot , \cdot ])$ and that of $(\mathcal{M}, \{ \cdot , \cdot \}_{M})$, where $\mathcal{M}$ stands for the Moyal algebra \cite{Compeans}.

Thus, within the WWM formalism the equation of motion (\ref{QEqMotion}) and the Gauss Law (\ref{QGaussLaw}) are written as
\begin{equation}
\frac{\partial^{2} \mathcal{A}_{i}}{\partial t^{2}} + \{\mathcal{A}_{j}, \{\mathcal{A}_{j}, \mathcal{A}_{i}\}_{M}\}_{M} + \frac{1}{2} C_{ijkl} \{\mathcal{A}_{j}, \{\mathcal{A}_{k}, \mathcal{A}_{l} \}_{M}\}_{M} = 0, \label{MEqMotion}
\end{equation}
and
\begin{equation}
\Big\{ \frac{\partial^{2} \mathcal{A}_{i}}{\partial t^{2}}, \mathcal{A}_{i}\Big\}_{M}. \label{MGaussLaw}
\end{equation}

As usual, we write down the Lagrangian and the Hamiltonian giving rise to equation (\ref{MEqMotion}) as follows
\begin{eqnarray}
\mathcal{L}^{(M)} &=& \frac{\partial \mathcal{A}_{i}}{\partial t} \star \frac{\partial \mathcal{A}_{i}}{\partial t} + \frac{1}{2}  \{\mathcal{A}_{i}, \mathcal{A}_{j}\}_{M} \star \{\mathcal{A}_{i}, \mathcal{A}_{j}\}_{M} - \frac{1}{4} C_{ijkl} \{\mathcal{A}_{i}, \mathcal{A}_{j}\}_{M} \star \{\mathcal{A}_{k}, \mathcal{A}_{l}\}_{M}, \nonumber \\
&& \\
\mathcal{E}^{(M)} &=& \frac{\partial \mathcal{A}_{i}}{\partial t} \star \frac{\partial \mathcal{A}_{i}}{\partial t} - \frac{1}{2}  \{\mathcal{A}_{i}, \mathcal{A}_{j}\}_{M} \star \{\mathcal{A}_{i}, \mathcal{A}_{j}\}_{M} + \frac{1}{4} C_{ijkl} \{\mathcal{A}_{i}, \mathcal{A}_{j}\}_{M} \star \{\mathcal{A}_{k}, \mathcal{A}_{l}\}_{M}. \nonumber \\
&&
\end{eqnarray}
Therefore equations (\ref{MEqMotion}) and (\ref{MGaussLaw}) come from the deformed version of the Nahm equations in seven dimensions
\begin{equation}
\frac{\partial \mathcal{A}_{i}}{\partial t} = \frac{1}{2} C_{ijk} \{ \mathcal{A}_{j}, \mathcal{A}_{k} \}_{M}.
\label{MNE}
\end{equation}
Thus we got the Moyal-Nahm equations worked out  at \cite{Baker:1999ru} from the WWM formalism.\\

It remains elusive to us the generalization to higher dimensions of the well-known equivalence between the large $N$-limit of a su$(N)$ gauge theoy and the gauge theory valued in the deformed algebra of diffeomorfisms of the symplectic surface $\Sigma$, $\mathfrak{sdiff}_{\hbar}(\Sigma)$. In \cite{Compean1}, the three dimensional Nahm's equations with gauge group $SU(\infty)$ provide us with three hamiltonian vector fields (justly as dealt in by Ward in \cite{wardlargeN}) in the limit $\hbar \rightarrow 0$. In the present case we get seven functions but we do not know if these would correspond to seven  hamiltonian vector fields defined over some higher dimensional phase space $\Gamma$ generalizing $\Sigma$\footnote{It would be extremely interesting to find some large $N$-limits of certain Lie algebras ${\cal G}_N$ such that we have in some sense
an equivalence ${\cal G}_\infty \cong \mathfrak{sdiff}_{\hbar}(\Gamma)$.}. In the present case on $\Sigma$ the complete Lagrangian and Hamiltonian are
\begin{equation}
L^{(M)} = -\int_{\Sigma} \dif p \dif q \; \mathcal{L}^{(M)}, \quad E^{(M)} = -\int_{\Sigma} \dif p \dif q \; \mathcal{E}^{(M)},
\end{equation}
with $p$ and $q$ being the local coordinates on $\Sigma$.

We will assume that the functions $\mathcal{A}_{i}(t, p, q; \hbar)$ are just a formal series of the deformation paremeter $\hbar$, that is
\begin{equation}
\mathcal{A}_{i}(t, p, q; \hbar) = \sum_{n=0}^\infty \hbar^{n} \mathcal{A}_{i,n}(t, p, q).
\end{equation}


\section{The Quest for Solutions}

As is the case for the three dimensional version, we are now interesting to find solutions of the Moyal-Nahm equations (\ref{MNE}). In Ref. \cite{Baker:1999ru} the authors found a class of solutions in terms of elliptic functions, working with the gauge group $\mbox{SU}(2)$, and taking as an ansatz a direct sum decomposition for any representation of Pauli matrices $\sigma \in \mbox{su}(2)$. In such a case the $A$'s are diagonal justly of the generators of  $\mbox{su}(2)$. This fact make possible to use the same method from \cite{Compean1} to find solutions to the Moyal-Nahm equations in three dimensions. In the presnt case we will consider two cases: $\mbox{su}(2)$ and $\mathfrak{sl}(2, \mathbb{R})$.

Let $\Phi: \mathcal{G} \rightarrow \widehat{\mathcal{I}}$ be the algebra homomorphism between the gauge Lie algebra $\mathcal{G}$ and the Lie algebra of operators $\widehat{\mathcal{I}}$. We already know that $A_{i}(t) \in \mathfrak{M} \otimes \mathcal{G}$, thus $\Phi$ implies that
\begin{equation}
\widehat{\mathcal{A}}_{i} = \widehat{\mathcal{A}}_{i} (t) \in \mathfrak{M} \otimes \widehat{\mathcal{I}}.
\end{equation}

Taking $\tau_{i}$, as the generators of the gauge Lie Algebra, where $i$ runs on the dimension of it, we write $\widehat{\mathcal{X}}_{i} := \Phi(\tau_{i})$ for the corresponding operator. By the WWM correspondence these operators correspond to functions defined on $\Sigma$ as previously considered, that is
\begin{equation}
\mathcal{X}_{i}(t, p, q; \hbar) = \int_{-\infty}^{\infty} \Big\langle q - \frac{1}{2}\xi \Big| \widehat{\mathcal{X}}_{i} \Big| q + \frac{1}{2}\xi \Big\rangle \exp \left( \frac{\im}{\hbar}\xi \cdot  p \right) \dif \xi, \label{WWM}
\end{equation}
and they should be a solution of the Moyal-Nahm equations. The WWM formalism we are going work out in this paper strongly requires to know the solutions of the Nahm's equations in seven dimensions with any finite dimensional Lie algebra. The solution found \cite{Baker:1999ru} in terms of a matrix representation using the su$(2)$ Lie algebra is quite convenient for us. Although we use these results our solutions are different of those obtained in \cite{Baker:1999ru}.

\subsection{The $\mbox{SU}(2)$ case}

 Starting with the solutions obtained  in Ref. \cite{Baker:1999ru}, which are given by the next matrices, when we consider the gauge group $\mbox{SU}(2)$, and taking into account that $\tau_{i} = (\im/2)\sigma_{i}$, where $\sigma_{i}$ are the Pauli matrices
\begin{eqnarray}
A_{1} = - \left(
\begin{array}{ccc}
f_{1}\tau_{3} & 0 & 0 \\
0 & g_{1}\tau_{3} & 0 \\
0 & 0 & h_{1}\tau_{3}
\end{array}
\right), &
A_{2} = - \left(
\begin{array}{ccc}
f_{2}\tau_{1} & 0 & 0 \\
0 & g_{2}\tau_{2} & 0 \\
0 & 0 & \im h_{2}\tau_{3}
\end{array}
\right), \nonumber \\
A_{3} = - \left(
\begin{array}{ccc}
f_{3}\tau_{2} & 0 & 0 \\
0 & \im g_{3}\tau_{2} & 0 \\
0 & 0 & h_{3}\tau_{3}
\end{array}
\right), &
A_{4} = - \left(
\begin{array}{ccc}
f_{4}\tau_{2} & 0 & 0 \\
0 & \im g_{4}\tau_{3} & 0 \\
0 & 0 & h_{4}\tau_{1}
\end{array}
\right), \nonumber \\
A_{5} = - \left(
\begin{array}{ccc}
\im f_{5}\tau_{2} & 0 & 0 \\
0 & g_{5}\tau_{3} & 0 \\
0 & 0 & h_{5}\tau_{2}
\end{array}
\right), &
A_{6} = - \left(
\begin{array}{ccc}
f_{6}\tau_{3} & 0 & 0 \\
0 & g_{6}\tau_{2} & 0 \\
0 & 0 & \im h_{6}\tau_{2}
\end{array}
\right), \nonumber \\
A_{7} = - \left(
\begin{array}{ccc}
\im f_{7}\tau_{3} & 0 & 0 \\
0 & g_{7}\tau_{1} & 0 \\
0 & 0 & h_{7}\tau_{2}
\end{array}
\right), &
\end{eqnarray}
where the functions $f$'s, $g$'s and $h$'s are given by
\begin{eqnarray}
f_{7} &=& K_{1} f_{6} = K_{1} M_{1} f_{1} = K_{1} M_{1} Q_{1} \mbox{sn}(q_{1} t + d_{1}), \nonumber \\
f_{5} &=& K_{1} f_{4} = K_{1} M_{1} f_{3} = -\im K_{1} M_{1} Q_{1} \mbox{cn}(q_{1} t + d_{1}), \nonumber \\
f_{2} &=& \im q_{1} \mbox{dn}(q_{1} t + d_{1}), \\
g_{4} &=& K_{2} g_{5} = K_{2} M_{2} g_{1} = K_{2} M_{2} Q_{2} \mbox{sn}(q_{2} t + d_{2}), \nonumber \\
g_{3} &=& K_{2} g_{2} = K_{2} M_{2} g_{6} = -\im K_{2} M_{2} Q_{2} \mbox{cn}(q_{2} t + d_{2}), \nonumber \\
g_{7} &=& \im q_{2} \mbox{dn}(q_{2} t + d_{2}), \\
h_{2} &=& K_{3} h_{3} = K_{3} M_{3} h_{1} = K_{3} M_{3} Q_{3} \mbox{sn}(q_{3} t + d_{3}), \nonumber \\
h_{6} &=& K_{3} h_{7} = K_{3} M_{3} h_{5} = -\im K_{3} M_{3} Q_{3} \mbox{cn}(q_{3} t + d_{3}), \nonumber \\
h_{4} &=& \im q_{3} \mbox{dn}(q_{3} t + d_{3}).
\end{eqnarray}
Here $K_{i}$, $M_{i}$, $Q_{i}$, $q_{i}$, $d_{i}$ ($i = 1, 2, 3$) are constants related by
\begin{equation}
k_{i} = \frac{Q_{i}}{q_{i}}\sqrt{1 + M_{i}^{2}(1 - K_{i}^{2})},
\end{equation}
and the $\mbox{sn}(x)$, $\mbox{cn}(x)$, $\mbox{dn}(x)$ are the Jacobi elliptic functions satisfying
\begin{equation}
\mbox{sn}^{2}(x) + \mbox{cn}^{2}(x) = 1, \quad \mbox{dn}^{2}(x) + k_i^{2}\mbox{sn}^{2}(x) = 1.
\end{equation}

Using the results obtained in \cite{Compeans}, we have
\begin{eqnarray}
\Phi(\tau_{1}) = \widehat{\mathcal{X}}_{1} &:=& \im \beta \widehat{q} + \frac{1}{2\hbar} (\widehat{q}^{2} - 1) \widehat{p}, \\
\Phi(\tau_{2}) = \widehat{\mathcal{X}}_{2} &:=& -\beta \widehat{q} + \frac{\im}{2\hbar} (\widehat{q}^{2} + 1) \widehat{p}, \\
\Phi(\tau_{3}) = \widehat{\mathcal{X}}_{3} &:=& -\im \beta \widehat{1} - \frac{1}{\hbar} \widehat{q} \widehat{p},
\end{eqnarray}
where $\beta$ is a real constant, $\widehat{q}$ and $\widehat{p}$ are the position and momentum operators respectively, and with this, we find what is the WWM counterpart of the solutions just written.

First, we note that the $A$'s are diagonal, and by WWM formalism, we just apply (\ref{WWM}) on the $\widehat{\mathcal{X}}_{i}$'s. Thus, after some calculations we get
\begin{eqnarray}
\mathcal{X}_{1}(p, q; \hbar) &=& \im \left( \beta - \frac{1}{2} \right) q - \frac{1}{2\hbar} (q^{2} - 1)p, \\
\mathcal{X}_{2}(p, q; \hbar) &=& - \left( \beta - \frac{1}{2} \right) q - \frac{\im}{2\hbar} (q^{2} + 1)p, \\
\mathcal{X}_{3}(p, q; \hbar) &=& -\im \left( \beta - \frac{1}{2} \right) + \frac{1}{\hbar} qp.
\end{eqnarray}

Therefore, we get a set of the complex matrices $\mathcal{A}_{i}(t, p, q; \hbar)$
\begin{eqnarray}
\mathcal{A}_{1} = -\im\hbar \left(
\begin{array}{ccc}
f_{1} \mathcal{X}_{3} & 0 & 0 \\
0 & g_{1} \mathcal{X}_{3} & 0 \\
0 & 0 & h_{1} \mathcal{X}_{3}
\end{array}
\right), &
\mathcal{A}_{2} = -\im\hbar \left(
\begin{array}{ccc}
f_{2} \mathcal{X}_{1} & 0 & 0 \\
0 & g_{2} \mathcal{X}_{2} & 0 \\
0 & 0 & \im h_{2} \mathcal{X}_{3}
\end{array}
\right), \nonumber \\
\mathcal{A}_{3} = -\im\hbar \left(
\begin{array}{ccc}
f_{3} \mathcal{X}_{2} & 0 & 0 \\
0 & \im g_{3} \mathcal{X}_{2} & 0 \\
0 & 0 & h_{3} \mathcal{X}_{3}
\end{array}
\right), &
\mathcal{A}_{4} = -\im\hbar \left(
\begin{array}{ccc}
f_{4} \mathcal{X}_{2} & 0 & 0 \\
0 & \im g_{4} \mathcal{X}_{3} & 0 \\
0 & 0 & h_{4} \mathcal{X}_{1}
\end{array}
\right), \nonumber \\
\mathcal{A}_{5} = -\im\hbar \left(
\begin{array}{ccc}
\im f_{5} \mathcal{X}_{2} & 0 & 0 \\
0 & g_{5} \mathcal{X}_{3} & 0 \\
0 & 0 & h_{5} \mathcal{X}_{2}
\end{array}
\right), &
\mathcal{A}_{6} = -\im\hbar \left(
\begin{array}{ccc}
f_{6} \mathcal{X}_{3} & 0 & 0 \\
0 & g_{6} \mathcal{X}_{2} & 0 \\
0 & 0 & \im h_{6} \mathcal{X}_{2}
\end{array}
\right), \nonumber \\
\mathcal{A}_{7} = -\im\hbar \left(
\begin{array}{ccc}
\im f_{7} \mathcal{X}_{3} & 0 & 0 \\
0 & g_{7} \mathcal{X}_{1} & 0 \\
0 & 0 & h_{7} \mathcal{X}_{2}
\end{array}
\right). &
\end{eqnarray}
These matrices are solutions of the Moyal-Nahm equations if indeed the ${\cal X}$'s satisfy
\begin{equation}
\{\mathcal{X}_{1}, \mathcal{X}_{2}\}_{M} = \frac{-1}{\im \hbar} \mathcal{X}_{3}, \quad
\{\mathcal{X}_{2}, \mathcal{X}_{3}\}_{M} = \frac{-1}{\im \hbar} \mathcal{X}_{1}, \quad
\{\mathcal{X}_{3}, \mathcal{X}_{1}\}_{M} = \frac{-1}{\im \hbar} \mathcal{X}_{2},
\end{equation}
which is precisely the case.

´
\subsection{The $\mbox{SL}(2, \mathbb{R})$ case}

 Now we would like to discuss the case when the gauge group is  $\mbox{SL}(2, \mathbb{R})$. First of all, the relation between the generators of the Lie algebra is
\begin{equation}
[\tau_{1}, \tau_{2}] = \tau_{3}, \quad [\tau_{2}, \tau_{3}] = \tau_{1}, \quad [\tau_{3}, \tau_{1}] = -\tau_{2}.
\end{equation}

Hence, the solutions for Nahm equations are no so different in form from that case of $\mbox{SU}(2)$, in this case we have
\begin{eqnarray}
A_{1} = - \left(
\begin{array}{ccc}
f'_{1}\tau_{3} & 0 & 0 \\
0 & g'_{1}\tau_{3} & 0 \\
0 & 0 & h'_{1}\tau_{3}
\end{array}
\right), &
A_{2} = - \left(
\begin{array}{ccc}
f'_{2}\tau_{1} & 0 & 0 \\
0 & g'_{2}\tau_{2} & 0 \\
0 & 0 & \im h'_{2}\tau_{3}
\end{array}
\right), \nonumber \\
A_{3} = - \left(
\begin{array}{ccc}
f'_{3}\tau_{2} & 0 & 0 \\
0 & \im g'_{3}\tau_{2} & 0 \\
0 & 0 & h'_{3}\tau_{3}
\end{array}
\right), &
A_{4} = - \left(
\begin{array}{ccc}
f'_{4}\tau_{2} & 0 & 0 \\
0 & \im g'_{4}\tau_{3} & 0 \\
0 & 0 & h'_{4}\tau_{1}
\end{array}
\right), \nonumber \\
A_{5} = - \left(
\begin{array}{ccc}
\im f'_{5}\tau_{2} & 0 & 0 \\
0 & g'_{5}\tau_{3} & 0 \\
0 & 0 & h'_{5}\tau_{2}
\end{array}
\right), &
A_{6} = - \left(
\begin{array}{ccc}
f'_{6}\tau_{3} & 0 & 0 \\
0 & g'_{6}\tau_{2} & 0 \\
0 & 0 & \im h'_{6}\tau_{2}
\end{array}
\right), \nonumber \\
A_{7} = - \left(
\begin{array}{ccc}
\im f'_{7}\tau_{3} & 0 & 0 \\
0 & g'_{7}\tau_{1} & 0 \\
0 & 0 & h'_{7}\tau_{2}
\end{array}
\right), &
\end{eqnarray}
with the functions appearing defined by
\begin{eqnarray}
f'_{7} &=& K_{1} f'_{6} = K_{1} M_{1} f'_{1} = \frac{\im}{k_{1}}K_{1} M_{1} Q_{1} \mbox{ns}(q_{1} t + d_{1}), \nonumber \\
f'_{5} &=& K_{1} f'_{4} = K_{1} M_{1} f'_{3} = \frac{1}{k_{1}} K_{1} M_{1} Q_{1} \mbox{cs}(q_{1} t + d_{1}), \nonumber \\
f'_{2} &=& \im q_{1} \mbox{ds}(q_{1} t + d_{1}), \\
g'_{4} &=& K_{2} g'_{5} = K_{2} M_{2} g'_{1} = \frac{\im}{k_{2}}K_{2} M_{2} Q_{2} \mbox{ns}(q_{2} t + d_{2}), \nonumber \\
g'_{3} &=& K_{2} g'_{2} = K_{2} M_{2} g'_{6} = \frac{1}{k_{2}} K_{2} M_{2} Q_{2} \mbox{cs}(q_{2} t + d_{2}), \nonumber \\
g'_{7} &=& \im q_{2} \mbox{ds}(q_{2} t + d_{2}), \\
h'_{2} &=& K_{3} h'_{3} = K_{3} M_{3} h'_{1} = \frac{\im}{k_{3}}K_{3} M_{3} Q_{3} \mbox{ns}(q_{3} t + d_{3}), \nonumber \\
h'_{6} &=& K_{3} h'_{7} = K_{3} M_{3} h'_{5} = \frac{1}{k_{3}} K_{3} M_{3} Q_{3} \mbox{cs}(q_{3} t + d_{3}), \nonumber \\
h'_{4} &=& \im q_{3} \mbox{ds}(q_{3} t + d_{3}).
\end{eqnarray}
Here again, $\mbox{ns}(x)$, $\mbox{cs}(x)$ and $\mbox{ds}(x)$ are elliptic functions.

Following the procedure previously applied, using the isomorphism $\Phi(\tau_{i}) = \widehat{\mathcal{X}}_{i}$ \cite{Compeans}, we have the unitary operator version of the algebra $\mathfrak{sl}(2, \mathbb{R})$
\begin{eqnarray}
\Phi(\tau_{1}) = \widehat{\mathcal{X}}_{1} &:=& \frac{\im}{4} \left(\frac{\widehat{p}^{2}}{\hbar^{2}} + \frac{\delta}{\widehat{q}^{2}} - \widehat{q}^{2} \right), \\
\Phi(\tau_{2}) = \widehat{\mathcal{X}}_{2} &:=& \frac{\im}{4} \left(\frac{\widehat{p}^{2}}{\hbar^{2}} + \frac{\delta}{\widehat{q}^{2}} + \widehat{q}^{2} \right), \\
\Phi(\tau_{3}) = \widehat{\mathcal{X}}_{3} &:=& \frac{\im}{2} \left(\frac{\widehat{q} \widehat{p}}{\hbar} - \frac{\im}{2} \right)
\end{eqnarray}
where $\delta$ is a constant.

WWM correspondence (\ref{WWM}) give the functions associated with these operators are
\begin{eqnarray}
\mathcal{X}_{1} &=& \frac{\im}{4} \left(\frac{p^{2}}{\hbar^{2}} + \frac{\delta}{q^{2}} - q^{2}\right), \\
\mathcal{X}_{2} &=& \frac{\im}{4} \left(\frac{p^{2}}{\hbar^{2}} + \frac{\delta}{q^{2}} + q^{2}\right), \\
\mathcal{X}_{3} &=& \frac{\im}{2\hbar} qp.
\end{eqnarray}

\indent Finally, the complex matrices solving the Moyal deformation of Nahm equations take the form
\begin{eqnarray}
\mathcal{A}_{1} = - \im\hbar \left(
\begin{array}{ccc}
f'_{1}\mathcal{X}_{3} & 0 & 0 \\
0 & g'_{1}\mathcal{X}_{3} & 0 \\
0 & 0 & h'_{1}\mathcal{X}_{3}
\end{array}
\right), &
\mathcal{A}_{2} = - \im\hbar \left(
\begin{array}{ccc}
f'_{2}\mathcal{X}_{1} & 0 & 0 \\
0 & g'_{2}\mathcal{X}_{2} & 0 \\
0 & 0 & \im h'_{2}\mathcal{X}_{3}
\end{array}
\right), \nonumber \\
\mathcal{A}_{3} = - \im\hbar \left(
\begin{array}{ccc}
f'_{3}\mathcal{X}_{2} & 0 & 0 \\
0 & \im g'_{3}\mathcal{X}_{2} & 0 \\
0 & 0 & h'_{3}\mathcal{X}_{3}
\end{array}
\right), &
\mathcal{A}_{4} = - \im\hbar \left(
\begin{array}{ccc}
f'_{4}\mathcal{X}_{2} & 0 & 0 \\
0 & \im g'_{4}\mathcal{X}_{3} & 0 \\
0 & 0 & h'_{4}\mathcal{X}_{1}
\end{array}
\right), \nonumber \\
\mathcal{A}_{5} = - \im\hbar \left(
\begin{array}{ccc}
\im f'_{5}\mathcal{X}_{2} & 0 & 0 \\
0 & g'_{5}\mathcal{X}_{3} & 0 \\
0 & 0 & h'_{5}\mathcal{X}_{2}
\end{array}
\right), &
\mathcal{A}_{6} = - \im\hbar \left(
\begin{array}{ccc}
f'_{6}\mathcal{X}_{3} & 0 & 0 \\
0 & g'_{6}\mathcal{X}_{2} & 0 \\
0 & 0 & \im h'_{6}\mathcal{X}_{2}
\end{array}
\right), \nonumber \\
\mathcal{A}_{7} = -\im\hbar \left(
\begin{array}{ccc}
\im f'_{7}\mathcal{X}_{3} & 0 & 0 \\
0 & g'_{7}\mathcal{X}_{1} & 0 \\
0 & 0 & h'_{7}\mathcal{X}_{2}
\end{array}
\right), &
\end{eqnarray}

\noindent provided that
\begin{equation}
\{\mathcal{X}_{1}, \mathcal{X}_{2}\}_{M} = \frac{1}{\im \hbar} \mathcal{X}_{3}, \quad \{\mathcal{X}_{2}, \mathcal{X}_{3}\}_{M} = \frac{1}{\im \hbar} \mathcal{X}_{1}, \quad \{\mathcal{X}_{3}, \mathcal{X}_{1}\}_{M} = \frac{-1}{\im \hbar} \mathcal{X}_{2}.
\end{equation}

\vskip 1truecm
\section{Final Remarks}

In this paper we derived the Moyal-Nahm equations in seven dimensions by using the WWM formalism. In order to find solutions of the deformation of Nahm equations we take advantage of the solutions found by Baker and Fairlie in Ref \cite{Baker:1999ru} for a finite dimensional Lie algebra in terms of matrix representations using the su(2) algebra. Although we use from these Lie algebra solution, however the solutions are different from those obtained in \cite{Baker:1999ru}. We also find solutions for the algebra $\mathfrak{sl}(2, \mathbb{R})$. We don't take the limit $\hbar \to 0$ to find the large $N$-limit, since at the present time it is not well understood how to relate the large-$N$ limit of some Lie algerbra ${\cal G}_N$ with $\mathfrak{sdiff}_{\hbar}(\Gamma)$ for higher dimensions of $\Gamma$ than 2. We would like to address this problem in a near future and compare systematically with the results of the proposal of Ref. \cite{yo}.  Another problem that we would like to address is that of the integrability of self-dual equations and their descendants through a version of the twistor transform in  higher dimensions. Finally we are interesting in studying the canonical structures as was discussed in Ref. \cite{Chakravarty:1991bt}, now in the context of the higher-dimensional version of the AJS equations proposed in \cite{yo}. Our aim here is to obtain an octonionic version of the Plaba\'nski heavenly equations \cite{Plebanski:1975wn}. Some of this work is in progress and will be reported elsewhere.

\vskip 1truecm
\centerline{\bf Acknowledgments}

The work of A.M.-M. was supported by a CONACyT graduate  fellowship.

\vskip 1truecm

\bibliography{octaviostrings}
\addcontentsline{toc}{section}{Bibliography}
\bibliographystyle{TitleAndArxiv}


\end{document}